\documentclass[preprint,showpacs,twoside]{revtex4}

\usepackage[english]{babel}
\usepackage{graphics, epsfig ,color} 
\usepackage{amsmath,amssymb} 
\usepackage{array,hhline} 
\usepackage{fancyhdr}
\usepackage{times}

\def\bse{\begin{subequations}}
\def\ese{\end{subequations}}

\begin{document}

\title{Injection locking of a semiconductor laser to a multi-frequency reference beam}%

\author{T. Yang}
\affiliation{Centre for Quantum Technologies, National University of
Singapore , 117543 Singapore, Singapore}
\author{K. Pandey}
\affiliation{Centre for Quantum Technologies, National University of
Singapore , 117543 Singapore, Singapore}
\author{M. Giudici}
\affiliation{Institut Non Lin\'eaire de Nice, Universit\'e de Nice
Sophia-Antipolis, CNRS, 06560 Valbonne, France}
\author{D. Wilkowski}
\affiliation{Centre for Quantum Technologies, National University of
Singapore , 117543 Singapore, Singapore} \affiliation{Institut Non
Lin\'eaire de Nice, Universit\'e de Nice Sophia-Antipolis, CNRS,
06560 Valbonne, France} \affiliation{PAP, School of Physical and
Mathematical Sciences, Nanyang Technological University, 637371
Singapore, Singapore}
\date{\today{}}
\begin{abstract}
Injection locking is a well known and commonly used method for
coherent light amplification. Usually injection locking is done with
a single-frequency seeding beam. In this work we show that injection
locking may also be achieved in the case of multi-frequency seeding
beam when slave laser provides sufficient frequency filtering. One
relevant parameter turns out to be the frequency detuning between
the free running slave laser and each injected frequency component.
Stable selective locking to a set of three components separated of
$1.2\,$GHz is obtained for (positive) detuning values between zero
and $1.5\,$GHz depending on seeding power (ranging from 10 to 150
microwatt). This result suggests that, using distinct slave lasers
for each line, a set of mutually coherent narrow-linewidth
high-power radiation modes can be obtained \end{abstract}

%

\maketitle

\section{Introduction}

\begin{figure}
\includegraphics[width=0.48\textwidth]{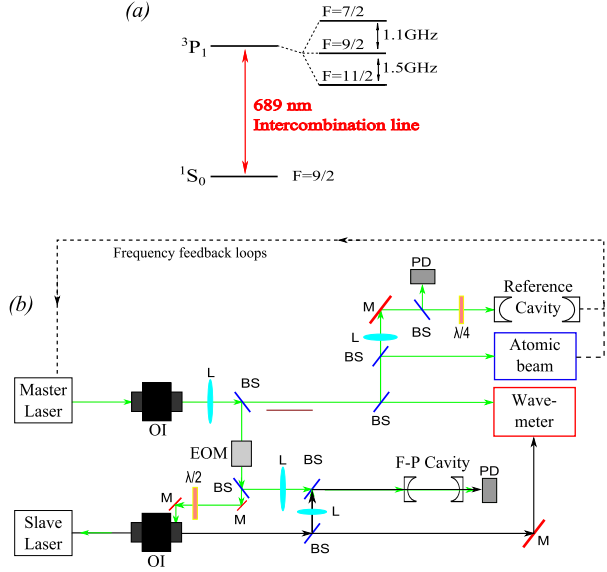}
\caption{ (a) Energy levels of the $^1S_0\!\!\rightarrow ^3\!\!\! P_1$ intercombination line of
$^{87}$Sr. The laser system address the Hyperfine
structure of the $^3 P_1$ state. (b) Schematic of the experimental
set-up. The master laser is frequency locked on a reference cavity and on an atomic beam.
The multi-frequency lines are generated in an EOM.
The frequency spectra of the master laser and the slave laser are analyzed trough a wavemeter and a Fabry-Perot cavity.
We use the following notation, L: Lens, M: Mirror, BS: Beam splitter, OI: Optical isolator and PD: Photodetector.}
\label{fig1}
\end{figure}

Laser optical injection consists in seeding a free running slave
laser with a master laser beam \cite{siegman86}. Since the laser is
nonlinear oscillator, this seeding leads a rich variety of dynamical
behaviors, including laser instabilities \cite{Tredicce:85}, route
to chaos \cite{Arecchi1984308}, excitability \cite{Kelleher:09},
extreme events \cite{PhysRevLett.107.053901}, and injection locking
\cite{stover:91}. Injection locking improves significantly the
performances of slave lasers and, in the case of semiconductor
lasers \cite{1071166}, it leads to several benefits for many
applications including optical communications, microwave signal
generation, all-optical signal processing, as illustrated in a
recent tutorial publication \cite{4838882}. In this paper we are
concerned by the transfer of the coherence properties of the master
oscillator toward an high-power slave laser, thus leading to strong
optical amplification of the reference oscillator output
\cite{Farinas:94}. In particular, we address the problem of
injection locking with a multi-frequency reference beam.
Our aim is to use each one of these frequency lines to injection
lock distinct power lasers, thus obtaining a set of mutually
coherent narrow-linewidth high-power radiation modes. This set of
coherent modes is very attractive for generating Raman laser
\cite{kasevich1992laser,ringot1999generation} and for coherently
controlling multiphotons absorbtion processes
\cite{mahmoudi2006light} but also for dense-wavelength-division
multiplexing \cite{917856} and millimiters wave generation
\cite{1263722}. In our case, we intend to use this injection locking
scheme to address the hyperfine levels of the $^{87}$Sr
$^3P_1$ state in the $1\,$GHz frequency domain (Fig. \ref{fig1}a).

Injection locking using multi-frequency master laser has been
analyzed recently using optical frequency combs as master radiation
\cite{moon:181110,ryu:141107,Moon:11}. The frequency comb is
generated with a femtosecond mode-locked Ti:sapphire laser or fiber
laser, while a single-mode distributed feedback semiconductor laser
is used as slave laser. The large number of modes and the small
mode-spacing ($100\,$MHz) emitted by the mode-locked laser needs to
be filtered before seeding the slave laser, otherwise it may prevent
the single-mode stability of injection locking
\cite{moon:181110,Lu:11}.

Our approach is based instead on a external cavity mounted tunable
single-mode semiconductor laser as master oscillator, sidebands are
obtained from the emitted radiation line  by using an Electro-Optic
Modulator (EOM)
\cite{szymaniec1997injection,shahriar2000demonstration}.  Under
specific conditions explained throughout the paper, we show that a
slave multi-quantum-well laser can be injection locked to a well defined
radiation line without being affected by the other
sidebands. Our scheme features several advantages with respect to
the mode-locked laser scheme discussed in
\cite{moon:181110,ryu:141107}: i) the frequency separation of the
lines can be varied simply adjusting the RF frequency of the EOM,
ii) the phase and the amplitude of each radiation line can be
independently controlled, iii) the carrier frequency can be
shifted quasi-continuously acting on the tunable diode laser, iv) filtering is simply obtained
by relying on the coupling with the slave laser resonator without
additional external filters and v) diode laser master oscillator is
more compact, flexible and cheaper than Ti:sapphire laser.

The paper is organized as follow: We first describe the
experimental set-up and the detection scheme of the lasers frequency
and power spectrum. Then we present and discuss in detail our
experimental results. In particular, we show that injection
locking only occurs when the frequency of the free running slave
laser is close to the frequency of the master oscillator.
Finally we draw our conclusion.\\

\section{Experimental set-up}
\label{Experimental set-up}

Master oscillator radiation is provided by a tunable external cavity
diode laser (Fig. \ref{fig1}b). This system is electronically locked
on an ultrastable reference external cavity using the
Pound-Drever-Hall technic \cite{drever1983laser}. Under this
condition, the frequency linewidth is below $1\,$kHz over one
second. The long term absolute frequency drift of the cavity is
below $30\,$Hz/s. it is corrected thanks to a saturated spectroscopy
set-up on Strontium atomic beam. The master laser is maintained
constant at an absolute frequency
$\nu_{\textrm{0}}$=$434848.50 (5)\,$GHz.

Before seeding, the master laser beam is sent through an EOM which
generates lateral bands at $\Omega=1.2\,$GHz. The phase modulation
index of the EOM is adjusted at $1.43$ such as the central band and
the $\pm 1$ lateral bands have the same amplitude. In this case
$90\%$ of the optical power is equally distributed among the three
sidebands, and we ignore the contribution of the higher order
sidebands.

After EOM, the master laser beam is sent into the slave laser using
a lateral port of an optical isolator (Fig. \ref{fig1}b). We control
the seeding power adjusting a $\lambda/2$ retardation wave plate in
front of the optical isolator port. The slave laser is a AlGaInP
multi-quantum-well semiconductor laser from Opnext lasing at
$0.69\,\mu$m (Model: HL6738MG). Its cavity has a free spectral range
of roughly $47\,$GHz.
The absolute frequencies of master and slave lasers are monitored
continuously by a lamdbameter with an accuracy of $60\,$MHz. The
frequency spectra of the lasers are monitored trough
the transmission of a scanning Fabry-Perot interferometer with a
free spectral range of $1.5\,$GHz.

\begin{figure}
\includegraphics[width=0.48\textwidth]{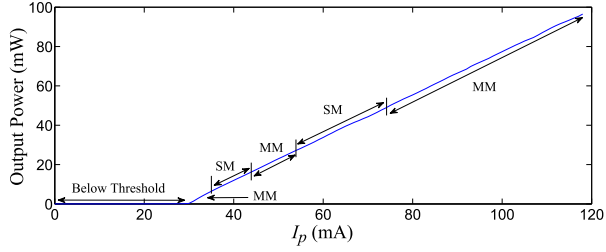}
\caption{L/I curve of the free running slave laser. Above the
threshold, SM means Single-mode laser emission, MM means
Multi-mode laser emission.} \label{fig2}
\end{figure}

The L/I curve of the free running slave laser is shown in Fig.
\ref{fig2}. The lasing threshold is at $I_p=30\,$mA. Increasing the pumping current $I_p$ further, we observe an
alternation of single-mode and multi-mode emission. This alternation
can be explained considering that, the temperature of the active region increases with the pumping
current. It induces a frequency shift of the gain curve together with a
variation of the refraction index, leading to a frequency shift of
the cavity resonances \cite{Petermann,1337016}. Because the cavity
resonances shift at a smaller rate with respect the gain
peak, an increase of the pumping current induces longitudinal mode
jumping towards higher wavelength. Multi-mode regimes occur when the
gain peak is not close enough to any longitudinal modes to impose
single-mode behavior, thus leading to multi-mode instability.

We set the pumping current of the slave laser in the range of
$I_p=55$-$75\,$mA where the laser is single-mode and mode
hop free. The frequency of the free running laser
$\nu_{\textrm{free}}$ is continuously and linearly tuned by changing
$I_p$ with a slope:
\begin{equation}
\partial\nu_{\textrm{free}}/\partial I_p=-3.85,\textrm{GHz/mA}. \label{slope}
\end{equation}
So in the current range explored, the laser output frequency has
shifted of more than 1.5 times its cavity free spectral range,
though maintaining single-mode behavior. At the injection current
value of $73\,$mA, the slave laser emission power is about $50\,$mW
and its frequency coincides with the master laser frequency.

\section{Single-frequency injection}

Semiconductor laser injection of a single-frequency has
been largely explored in semiconductor lasers both theoretically and
experimentally. Our observations do not differ from
what have been reported so far (see for example:
\cite{1071632,108,1072760,Wieczorek20051,1355-5111-9-5-001}).
However it worths to analyze the behavior of our slave laser in this
regime so we can pinpoint similarities and differences of the
multi-frequency injection discussed in the next section.

We did a systematic search of the injection locking domains as a
function of the seeding power and frequency detuning,
$\Delta\nu=\nu_{\textrm{free}}-\nu_{\textrm{0}}$, between the free
running slave laser and the master laser. $\Delta\nu$ is scanned
changing the pumping current of the slave laser. We limit our search
in the interval $I_p$ above defined, where $\Delta\nu$ can
be clearly defined.

\begin{figure}
\includegraphics[width=0.48\textwidth]{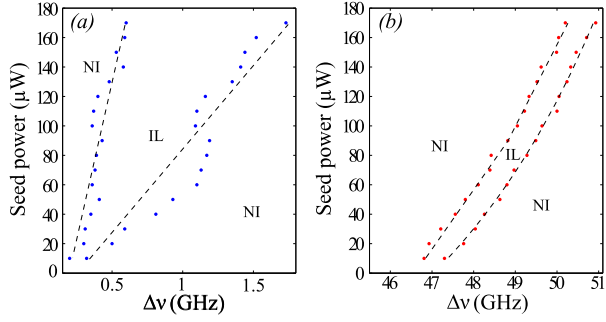}
\caption{Single-frequency injection locking domain of slave laser in
the frequency detuning-seeding power plane. The dots are
the measured values whereas the dash lines are not fits, but guide
the reader eyes. \emph{(a)}: injection locking domain when the free
running laser frequency is close to the seeding laser frequency
$\Delta\nu\approx 0$. \emph{(b)}: injection locking domain when the
free running laser frequency is detuned by a free spectral range the
seeding laser frequency $\Delta\nu\approx 47\,$GHz  IL (NI) means
(no) injection locking on the master oscillator. The detuning values refer to the master frequency
of $\nu_0=434848.50(5),$ GHz} \label{fig3}
\end{figure}

The results are summarized in Fig. \ref{fig3}. In the graph,
the dots materialize the boundary between the injection locked
domains and regions where the slave laser is not interacting with
the seeding beam or where injection leads to a non single-frequency
emission. Injection locking is observed for a minimum seeding power
of about $10\,\mu$W corresponding to an amplification ratio of the
laser radiation of $50\,$dB.

For the lowest seeding power values, injection locking occurs when
the free running slave frequency $\nu_{\textrm{free}}$ and the
master laser frequency $\nu_{\textrm{0}}$ almost coincide (Fig.
\ref{fig3}a). More precisely, the lower limit of frequency
difference for injection locking is $\Delta\nu\simeq +0.2\,$GHz.
This red/blue asymmetry of the injection frequency with respect to the free
running frequency is well known \cite{108,1072760}. It can be easily
understood considering that light injection inside the slave
laser decreases the carrier number in the active region, thus
affecting the refraction index and shifting the cavity resonances of
the slave laser towards lower frequencies. In order to compensate
this effect it is necessary to place the slave laser emission line
(which coincide with one cavity resonance) on the blue side with
respect the master frequency. The required positive detuning value
increases with the injected power and it results in a progressive
shift towards higher detuning values of the injection locking
region. We note as well that the injection locking region get
broader when the seeding power is increased. The
maximum seeding power used in this work is $180\,\mu$W, however we
observed that injection locking persists at even higher
seeding power.

Increasing $\Delta\nu$ (by decreasing slave pumping current) beyond
the high frequency borderline of Fig. \ref{fig3}a, the injection
locking is lost. It is recovered, when the detuning corresponds
roughly to the free spectral range of the slave laser such as the neighbor longitudinal mode gets almost resonant with the seeding
beam frequency. We obtain then the locking region depicted in Fig.
\ref{fig3}b. As for the previous case, we observed a blue shift of
the free running laser frequency for the injection locking condition
when the seed power is increased. However the injection locking
frequency range is
now narrower. This narrowing is probably due to a more stringent mode competition between the injection locked mode and the free running lasing mode.\\

\section{Multi-frequency seeding}

It has been shown that a multi-frequency injection affects the stability of injection locking
\cite{Lu:11}. This fundamental problem can be solved by using a
slave laser whose resonator is capable of selecting a single
frequency component and filtering out the remaining ones. In other
words, slave laser frequency selectivity is the key point for
achieving stable injection locking in presence of a multi-frequency
master radiation. Accordingly, slave laser selectivity determines a
minimum frequency separation between the master components for which
injection locking can still be achieved.

Let us analyze injection locking of our slave laser to a
multi-frequency master radiation obtained by using an EOM to
generate sidebands on the master beam, as described in Section
\ref{Experimental set-up}

\begin{figure}
\includegraphics[width=0.48\textwidth]{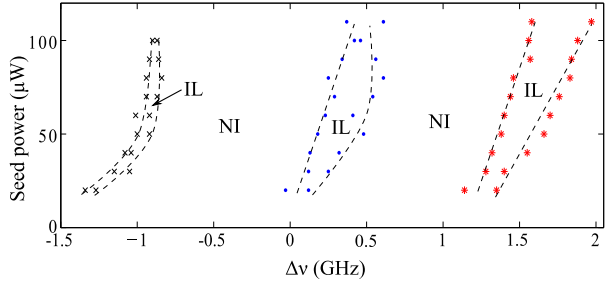}
\caption{Multi-frequency injection locking domains of slave laser in
the frequency detuning-seeding power plane. The ordinate axis gives
the total seeding power. The detuning is measured with respect the
central frequency of the injected beam. The slave laser is injected
by three lines separated by $1.2\,$GHz. The dots, crosses and stars
are the measured values whereas the dash lines are not fits, but
guide the reader eyes. IL (NI) means (no) injection locking on the master
oscillator. The detuning
values refer to the master frequency of $\nu_0=434848.50(5),$
GHz} \label{fig4}
\end{figure}

\begin{figure}
\includegraphics[width=0.5\textwidth]{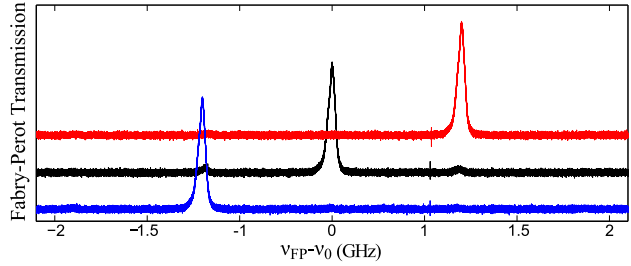}
\caption{Fabry-Perot transmission spectra of slave laser injected
by three lines separated by $1.2\,$GHz for a seeding power of $50\,\mu$W. The blue, dark and red
curves correspond respectively to injection locking of the red,
central, and blue line. The spectra are obtained for different
$\Delta\nu$ values within the injection locking domains depicted in
Fig. \ref{fig4}. Spectra baselines have been shifted along the ordinate axis to
easily observed the presence of minor peaks. } \label{fig5}
\end{figure}

As for the single-frequency case, we scan the free running frequency
$\nu_{\textrm{free}}$ around the central frequency of the seeding
beam and we observe that the slave laser can be injection locked to
each frequency component of the master beam. Our
result are summarized in Fig. \ref{fig4}, showing the locking ranges
and in Fig. \ref{fig5}, showing the optical spectrum emission of the
injection-locked slave laser. Within the injection locking domains,
more than $90\%$ of the slave power is in the selected sideband. As it occurs in the single-frequency
case, we observe that the detuning range where injection locking
occurs is asymmetrically shifted towards positive values of
$\Delta\nu$ (Fig. \ref{fig4}). This (positive) detuning
range remains however smaller than the separation of the frequency
components of the master beam. Injection locking is achieved for
seeding powers smaller than $110\,\mu$W. Instead for seeding power
larger than $110\,\mu$W the filtering of the slave laser resonator
is not anymore sufficient to avoid competition between the injected
frequency lines and the system becomes unstable.

We notice in Fig. \ref{fig4} that the injection locking range of the
blue sideband is larger than the injection locking range of
the red side band. This feature is not universal and different
values of the central frequency of the seeding laser leads to different shape of the injection locking domains.

If we shift the free running slave laser frequency up to one free
spectral range, as we did for the single-frequency case, selective
injection locking is not achieved anymore. More precisely, not more
than $60\%$ of the slave power is in one specific line and the
system is usually dynamically unstable.

We can then conclude that selective injection locking occurs for a
multi-frequency seeding beam with a band spacing of
$\Omega=1.2\,$GHz, provided that the free running slave laser is
tuned close to the central injected frequency, i.e. $\Delta\nu\simeq
0$. Similar studies, but with a larger band spacing of $\Omega\sim
10\,$GHz have shown that selective injection locking is obtained
without mentioning the important role $\Delta\nu$ observed in this current
work
\cite{ringot1999generation,szymaniec1997injection,shahriar2000demonstration}.

\begin{figure}
\includegraphics[width=0.50\textwidth]{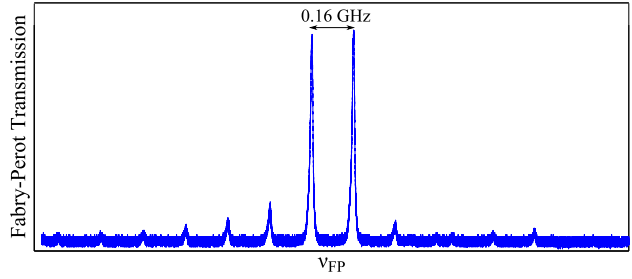}
\caption{Example of a Fabry-Perot transmission spectrum of slave
laser injected by two lines separated by $0.16\,$GHz for a seeding power of $50\,\mu$W.
The asymmetry in the spectrum
is probably due to an imbalance of the injection power of the two
lines which, in this particular case, cannot be accurately controlled.
\label{fig6}}
\end{figure}

By slightly modifying the set-up described in Section
\ref{Experimental set-up}, we test the multi-frequencies injection
with a smaller separation between the components. In this new set-up
the master beam is composed by only two frequency components whose
separation is of $\Omega=0.16\,$GHz. In this situation we were not
able to achieve injection locking, as also reported in similar
experiment \cite{moon:181110,ryu:141107}. In Fig. \ref{fig6} we plot
the optical spectrum of the slave laser emission when injected and
it is evident that both lines are present. Similar spectra where
obtained by changing the value of $\Delta\nu$, suggesting that
frequency selectivity of the slave resonator is not strong enough to
filter a single injection band. Due to nonlinear mixing of the two
lines in the gain medium, the slave laser spectrum contains also
other lateral bands appearing at frequencies separated from the two
main lines by an interval $\pm n\Omega$, where $n$ is an integer
(Fig. \ref{fig6}). The amplitude of these extra lateral bands
depends on the seeding conditions.

\section{Conclusion}

In this paper, we study the injection properties of a multi-quantum
well semiconductor laser subject to multi-frequency seeding. In
particular we address the important issue of injection locking of
the laser to a single band, relying on the frequency
selectivity of the slave laser only. Using a frequency spacing of
the seeding sideband of $1.2\,$GHz, single band injection locking
occurs only if the free running slave laser is close to the
injected frequency. If the slave laser is detuned by one frequency
spectral range, single band injection locking is lost. This result
shows that the frequency selectivity to injection of the slave laser
seems to crucially depend on the frequency difference between the
slave laser and the seeding laser. We also observe that for a
frequency spacing of the seeding sidebands of $0.16\,$GHz, single
line injection locking is not possible. Under this
condition the injected slave laser always exhibits a multi-frequency
spectrum.

\bibliographystyle{unsrt}
\bibliography{Injection_max}

\end{document}